\documentclass[twocolumn,showpacs,preprintnumbers,amsmath,amssymb,prl,superscriptaddress]{revtex4}

\usepackage{amsmath,amssymb,amsthm,color}
\usepackage{graphicx}
\usepackage{dcolumn}

\newcommand {\beq}{\begin{eqnarray}}
\newcommand {\eeq}{\end{eqnarray}}

\begin{document}

\preprint{IPMU11-0198}

\title{Gravity Dual for Hofman-Strominger Theorem}

\author{Yu Nakayama}

\affiliation{Institute for the Physics and Mathematics of the Universe,  \\ Todai Institutes for Advanced Study,
University of Tokyo, \\ 
5-1-5 Kashiwanoha, Kashiwa, Chiba 277-8583, Japan}

\affiliation{California Institute of Technology, 452-48, Pasadena, California 91125, USA}


\begin{abstract}
We provide a gravity counterpart of the theorem by Hofman and Strominger that in $(1+1)$ dimension, chiral scale invariance indicates chiral conformal invariance. We show that the strict null energy condition gives a sufficient condition to guarantee the symmetry enhancement. We also investigate a possibility to construct holographic c-function that decreases along the holographic renormalization group flow.
\end{abstract}

\maketitle

\section{Introduction}
Non-trivial symmetry enhancement is mysteriously common in classical general relativity. The oldest example would be the Israel theorem \cite{Israel} that states that any static asymptotically flat black hole must be spherically symmetric. This kind of uniqueness theorem  has played an important role in understanding  the holographic nature of the quantum gravitational theory.

On the other hand, we encounter symmetry enhancement in quantum field theories as well. The particular example we will be interested in this paper is the one proved by Zamolodchikov and Polchinski \cite{Zamolodchikov:1986gt}\cite{Polchinski:1987dy} (see also \cite{Mack1}) that states in (1+1) dimensional unitary relativistic quantum field theories, the scale invariance is automatically enhanced to the full conformal invariance. While its naive generalization in $(d+1)$ dimensional relativistic quantum field theories with $d>3$ has a counterexample \cite{Jackiw:2011vz}\cite{ElShowk:2011gz}, the situation in $d=2$ and $d=3$ is still open and has attracted renewed attention \cite{Dorigoni:2009ra}\cite{Antoniadis:2011gn}\cite{Fortin:2011ks}\cite{Fortin:2011sz}.

In our series of works \cite{Nakayama:2009qu}\cite{Nakayama:2009fe}\cite{Nakayama:2010wx}\cite{Nakayama:2011zw}, we have considered the gravity counterpart of the argument by Zamolodchikov and Polchinski, with possible higher dimensional generalizations as well. We have shown that the symmetry of the metric is always enhanced to the full $AdS_{d+2}$ group when we assume the isometry of $(d+1)$ dimensional Poincar\'e group and scale transformation. Furthermore, the additional null energy condition on the matter will forbid the violation of the $AdS_{d+2}$ isometry by non-trivial matter condensation. The holographic argument suggests a deep connection to the equivalence between scale invariance and conformal invariance and the c-theorem \cite{Zamolodchikov:1986gt} (or a-theorem \cite{Komargodski:2011vj}\cite{Nakayama:2011wq} in (1+3) dimension) that claims the existence of the monotonically decreasing function along the renormalization group flow in relativistic field theories.

Recently, Hofman and Strominger generalized the theorem by Zamolodchikov and Polchinski with less assumed symmetry \cite{Hofman:2011zj}. They studied (1+1) dimensional local quantum field  theories with only a chiral global scaling symmetry without assuming Lorentz invariance. They showed that the enhanced chiral conformal symmetry always follows under some technical assumptions. In this paper, we would like to study the gravity counterpart of the statement. We show that the strict null energy condition gives a sufficient condition to guarantee the symmetry enhancement in the holographic dual.

\section{Field theory argument}
The chiral scale invariance studied in \cite{Hofman:2011zj} states that the theory is invariant under the translation
\begin{align}
 t \to t + \epsilon_t \ , \ \ x \to x + \epsilon_x \ ,
\end{align}
and the chiral dilatation
\begin{align}
 t \to \lambda t \ 
\end{align}
in $(1+1)$ dimensional local quantum field theories.
Correspondingly, the theory possesses three conserved charges $H$, $P$, and $D$ with the commutation relation
\begin{align}
i[D,H] = H \ , \ \ i[D,P] = 0\ , \ \ i[H,P] = 0 \ .
\end{align}
All the symmetries must be linearly realized in a unitary manner.

We assume that the spectrum of the dilatation operator is diagonalizable and discrete. Note that the diagonalizability of the dilatation is not obvious when the theory is not conformal invariant: the internal rotation might give an antisymmetric part in the dilatation matrix \cite{Fortin:2011sz}.
We also assume that the theory has a local description so that the corresponding symmetry currents are all well-defined. Again this is not obvious because the condition is violated in the scale but non-conformal field theory studied in  \cite{Jackiw:2011vz}\cite{ElShowk:2011gz}.

From the translational invariance and the locality, the theory possesses a conserved energy-momentum tensor
\begin{align}
\partial_x T_{tx} + \partial_t T_{xx} = 0 \ , \ \ \partial_x T_{tt} + \partial_t T_{xt} = 0 \ ,
\end{align}
which is not necessarily symmetric $T_{xt} \neq T_{tx}$ due to the lack of Lorentz invariance.
The chiral scale invariance implies that the ``trace" of the energy-momentum tensor must be  given by the ``divergence" of the ``virial current":
\begin{align}
T_{xt} = \partial_t J_x + \partial_x J_t \ .
\end{align}
Then the chiral dilatation current 
\begin{align}
D_t = t T_{tt} - J_t \ , \ \ D_x = t T_{xt} - J_x \ 
\end{align}
is conserved: $\partial_x D_t + \partial_t D_x = 0$.

As discussed in  \cite{Hofman:2011zj}, we can always remove $J_t$ by defining the new conserved energy-momentum tensor
\begin{align}
\tilde{T}_{tt} = T_{tt} + \partial_t J_t \ , \ \ \tilde{T}_{xt} = T_{xt} - \partial_x J_t \ , \label{impr}
\end{align}
which is still conserved. When, in addition, $\partial_tJ_x$ vanishes, the theory possesses the chiral special conformal transformation induced by the conserved current
\begin{align}
K_t = t^2 \tilde{T}_{tt} \ , \ \ K_x = 0 \ 
\end{align}
together with the infinite tower of the chiral Virasoro symmetry ($L^n_t = t^{n} \tilde{T}_{tt}, \ L^n_x = 0$). The chiral special conformal transformation $K$ with the chiral dilatation will generate the $SL(2) \times U(1)$ subalgebra
\begin{align}
i[K,H] = D \ , \ \ i[D,K] = -K \ , \ \ i[K,P] = 0 \ .
\end{align}

The vanishing of $\partial_t J_x$ in unitary quantum field theories comes from the fact that the chiral scale invariance demands $\langle J_x(x,t) J_x(0) \rangle = f(x)$, indicating $\partial_t J_x = 0$ from the unitarity  \cite{Hofman:2011zj} (or more precisely {\it if} the analogue of the Reeh-Schlieder theorem \cite{RS} is true: in relativistic field theories, the proof requires the microscopic causality in addition to the unitarity).  
This shows that the chiral scale invariant field theories in (1+1) dimension are automatically invariant under the full chiral conformal transformation (with various technical assumptions).

Among the assumptions, the equivalence of the Reeh-Schlieder theorem, which the authors of \cite{Hofman:2011zj} used with no justification, was crucial. Since the Lorentz invariance is claimed to be abandoned, we do not know whether the assumption of the microscopic causality has its physical origin. Relatedly, while it is asserted that the translational symmetry $P$ did not play any important role in the emergence of the chiral special conformal invariance, it {\it did} play a significant role. For instance, if we compactified the $x$ direction, the microscopic causality might be lost. The Schr\"odinger field theory is an example of such a setup, where the Reeh-Schlieder theorem is not true, although possibly with some other reasons the symmetry enhancement may occur.
 In the following, we will discuss the gravity counterpart of this symmetry enhancement.

\section{Pure gravity}
Let us consider the $(1+2)$ dimensional metric that will exhibit the chiral scale invariance as an isometry. At least locally, we can safely assume that the isometry corresponding to the chiral scale invariance is realized by the translation
\begin{align}
 t \to t + \epsilon_t \ ,  \ \ x \to x + \epsilon_x
\end{align}
and the chiral scale transformation
\begin{align}
 z \to \lambda z \ , \ \  t \to \lambda t\ ,  \ \ x \to x \ .
\end{align}

Under this transformation, the most generic chiral scale invariant metric is given by
\begin{align}
ds^2 &= -a \frac{dt^2}{z^2} - 2b\frac{dt dx}{z} + c dx^2 + e \frac{dz^2}{z^2} \cr
  &+   2p \frac{dt dz}{z^2} + 2q \frac{dx dz}{z} \ . \label{gene}
\end{align}
It is easy to observe that by a simple coordinate transformation $t \to t + sz$  and $x \to x + r\log z$, which is compatible with the assumed isometry, one can always remove $p$ and $q$, so we will set $p = q=0$ in the following.

The resulting metric actually possesses the enhanced $SL(2)\times U(1)$ isometry: 
\begin{align}
ds^2 &=  -a \frac{dt^2}{z^2} - 2b\frac{dt dx}{z} + c dx^2   + e\frac{dz^2}{z^2}  \cr
 &= \frac{1}{c} \left(\frac{b dt}{z} - c dx\right)^2 + \frac{e (dz)^2 - (a + \frac{b^2}{c}) dt^2 }{z^2}  \label{wads}
\end{align}
where the metric is invariant under the additional isometry
\begin{align}
\delta z  &= 2\epsilon \left(a+\frac{b^2}{c}\right)tz \ , \ \  \delta t = \epsilon \left(a+\frac{b^2}{c}\right) t^2 + \epsilon e z^2 \cr
\delta x &= \frac{2\epsilon be}{c} \log z \label{special}
\end{align}
which gives the chiral special conformal transformation. 
When $b =0$, the geometry is locally $AdS_2 \times R$.
When $c = 0$, it is locally given by the so-called null warped $AdS_3$ space that has the $SL(2) \times U(1)$ isometry.
We can reach  the maximally symmetric $AdS_3$ space-time, whose isometry is further enhanced to $SO(2,2)$ by taking the additional limit $a = c=0$.

The general metric \eqref{wads} is known as the warped $AdS_3$ space in the Poincar\'e coordinate. Our claim is that the chiral scale invariant metric in $(1+2)$ dimensional space-time is always isometric to the warped $AdS_3$ space. Due to the enhanced isometry \eqref{special}, the symmetry algebra of the warped $AdS_3$ is precisely that of the chiral conformal symmetry discussed in the previous section.
It is known that the warped $AdS_3$ metric hosts the additional infinite number of symmetries as an asymptotic symmetry group \cite{arXiv:0808.1911}\cite{arXiv:0906.1243}\cite{arXiv:1108.6091}\cite{arXiv:1109.0544}. Thus, for the purely geometric part, the symmetry group will be further extended to the chiral Virasoro symmetry. Our focus, however, is the classical part of the symmetry that is realized by the isometry of the space-time, so we will not pursue the emergence of the chiral Virasoro algebra further in this paper.

If we restrict ourselves to the classical symmetry realized by the isometry, our argument in this section, therefore, is completely consistent with the field theory analysis by  \cite{Hofman:2011zj} reviewed in the last section. As we have seen there, the obstruction to the enhanced special conformal symmetry is due to the existence of the non-trivial virial current, so if we only talk about metric degrees of freedom in the bulk and assume that the bulk-field/boundary-operator correspondence works here, all we have in the dual field theory will be the energy-momentum tensor alone and we cannot introduce any such obstruction. In the next section, we consider the situation with matter where holographic dual candidates for the non-trivial virial current appear.

\section{Matter contribution and strict null energy theorem}
To make the game non-trivial and more interesting, we may want to introduce the matter contribution. As we will see, after all, the warped $AdS_3$ metric does not solve the vacuum Einstein equation with a cosmological constant, so we need some matter contribution or a modification of the Einstein equation to support the geometry. 

For concreteness, let us consider the vector (1-form) matter field $A = A_{\mu} dx^\mu$, anticipating that it  will be dual to the  virial current of the boundary theory. The field condensation that is compatible with the chiral scale invariance will be
\begin{align}
A = \alpha \frac{dz}{z} + \beta {dx} + \gamma \frac{dt}{z} \ . \label{vector}
\end{align}

We observe that the only one linear combination of $(\alpha,\beta,\gamma)$ gives the 1-form which is also invariant under the chiral special conformal transformation \eqref{special}. Therefore, for each bulk vector field (without assuming any equation of motion), we seem to have two independent obstructions to obtain the enhanced special conformal isometry in the bulk field configuration. This corresponds to the fact that for each (non-conserved) current, we may construct non-trivial candidates for the virial current component $J_t$ and $J_x$, which can be taken independently. Note that when the current is conserved, the dual bulk field will become a gauge field, so we can always adjust $\alpha$ and $\beta$ by gauge transformation so that \eqref{vector} is invariant under the chiral special conformal transformation.

In \cite{Hofman:2011zj}, it was argued that one of the components of the virial current can be removed by a redefinition of the operator (see the discussion around \eqref{impr}), and we would like to show its counterpart in the  dual bulk theory side. The basic idea is that we redefine the metric by
\begin{align}
\tilde{g}_{\mu\nu} = g_{\mu\nu} + h A_{\mu} A_{\nu} \ .
\end{align}
A similar metric redefinition was studied by \cite{arXiv:1111.6978} in a related study of the holographic energy-momentum tensor.
The new metric is still invariant under the chiral scale transformation, but this redefinition reinstate the ``off-diagonal component" $p$ and $q$ in \eqref{gene} (when $\beta$ and $\gamma$ are non-zero). We can perform the coordinate transformation  $t \to t + s(h) z$  and $x \to x + r(h) \log z$ to go back to the canonical form of the warped $AdS_3$ metric. The point is that the last coordinate transformation will change the coefficient $\alpha$ in the vector condensation: $\tilde{\alpha} = \alpha + \beta r(h) + \gamma s(h) $. In particular, it allows us to remove one extra parameter in the vector condensation so that apart from the chiral conformal invariant combination (that is not affected by this procedure), we only have one parameter left in the vector condensation that would violate the chiral special conformal transformation in agreement with the field theory argument. This is in complete agreement with the possibility to remove $J_t$ in the field theory by a redefinition of the energy-momentum tensor.

Yet we are not able to remove the remaining vector condensation that corresponds to non-zero $J_x$ without imposing equations of motion. Our argument here, much like the discussions in the full conformal case studied in \cite{Nakayama:2010wx}, is that the introduction of the strict null energy condition is sufficient to remove this vector condensation. However, since the warped $AdS_3$ space is not maximally symmetric space, the discussion will become slightly more involved.

Let us compute the relevant quantity in the null energy condition $R_{\mu\nu} k^{\mu} k^\nu$ for any null vector $k^\mu$ with components $(zk^t, k^x, z)$ that satisfies $-a (k^t)^2 - 2 b (k^t k^x) + c (k^x)^2 + e = 0$ in the warped $AdS_3$ space-time. The combination is always semi-positive definite:
\begin{align}
R_{\mu\nu} k^{\mu} k^{\nu} = \frac{a}{e(b^2 + ac)} (b k^t - c k^x)^2 \ge 0 \ ,
\end{align}
indicating that the warped $AdS_3$ space is consistent with the null energy condition.
It is non-zero because the warped $AdS_3$ is not a solution of the vacuum Einstein equation with a cosmological constant. The non-zero curvature may be supported by the vector condensation or by the gravitational Chern-Simons term in the modification of the Einstein equation (topological massive gravity).

Now, the strict null energy condition relevant for us states that if there exists a null vector which saturates the energy inequality, where it is explicitly given  by $(k^t, k^x, k^z) = (b^{-1}, c^{-1}, \sqrt{ab^{-2}e^{-1} + c^{-1}e^{-1}})$, the matter configuration must be trivial i.e. invariant under all the isometry of the metric. In this case, while it is allowed to introduce the condensation which is invariant under the chiral special conformal transformation, it is forbidden to introduce the other condensation that is not invariant under the full warped $AdS_3$ isometry. Note that once we impose the strict null energy condition, not only the vector condensation but also all the other higher spin condensation that would possibly violate the special conformal invariance are forbidden. 

The strict null energy condition utilized here may sound slightly stronger than the one discussed in \cite{Nakayama:2010wx}. For instance, we have to specify the particular null vector to saturate the bound and make the meaning of ``trivial" more precise here. These specifications  were largely irrelevant in the full conformal case due to the large space-time symmetry of the $AdS_{d+2}$ space. For instance, the warped $AdS_3$ space must be supported by a non-trivial matter configuration, but we have to state the energy condition concisely and consistently so that the matter supporting the space-time is still allowed. The sufficient condition presented in the last paragraph establishes this subtle balance.

We claim that the strict null energy condition is more or less equivalent to the unitarity condition (together with the microscopic causality) on the dual field theory that was utilized to remove the virial current in \cite{Hofman:2011zj}. This is supported by the consideration of the black hole holography where the null energy condition gives a sufficient condition for the area non-decreasing theorem of the black hole horizon \cite{Hawking}. There, the strict condition demands that when the entropy stays the same, non-trivial things will never happen. In other words, ``zero-energy" state will not carry any information. This amounts to the unitarity condition used by \cite{Hofman:2011zj} that excludes the vacuum degeneracy.

\section{Monotonically decreasing function?}
Another interesting application of the holographic approach is to study the properties of the renormalization group flow of the field theories by studying the holographic renormalization group flow in the gravity side \cite{Girardello:1998pd}\cite{Freedman:1999gp}\cite{Myers:2010xs}. The holographic renormalization of the non-Lorentz invariant system  has not been scrutinized completely, but we may still derive some interesting inequalities purely from the gravitational argument. 

For instance, we allow the $z$ dependent warp factor in the warped $AdS_3$ metric:
\begin{align}
ds^2 = -a(z) \frac{dt^2}{z^2} - 2b(z)\frac{dt dx}{z} + c(z) dx^2 + e(z) \frac{dz^2}{z^2} \label{ans}
\end{align}
and compute the left hand side of the null energy condition $R_{\mu\nu} k^{\mu}
k^\nu$, which must be semi-positive definite, to constrain the possible $z$ dependence of the warp factors. The trick here is to choose the null vector so that at the fixed point, the equality is saturated. In our case, the relevant null vector  will be $(k^t, k^x, k^z) = (b^{-1}, c^{-1}, \sqrt{ab^{-2}e^{-1} + c^{-1}e^{-1}})$. Then, we obtain the strongest inequality which governs the flow of the warp factors. Actually, we can show that the null energy condition does not give us any information unless $c(z)$ is non-constant. When $c(z)$ is constant, $R_{\mu\nu} k^{\mu} k^\nu$ simply vanishes for the particular null vector.

At this point, it is not clear to us whether the inequality obtained in this way is integrable or not for the most generic flow. Besides, when the flow is not {\it manifestly} invariant under the parity, we might have to introduce $p$ and $q$ again, which will complicate the analysis even further, while at the fixed point we may remove $p$ and $q$ by a coordinate transformation.

 For illustration, let us look at a simple flow, where the $z$ dependence appears only in the overall factor:
\begin{align}
ds^2 = f(z) \left(-a \frac{dt^2}{z^2} - 2b\frac{dt dx}{z} + c dx^2 + e \frac{dz^2}{z^2}  \right)
\end{align}
 then we can integrate the null energy condition so that
\begin{align}
R_{\mu\nu} k^{\mu}k^{\nu} &= 3z^2 (f')^2 f^{-3} - 4 z f' f^{-2} - 2z^2 f^{-2} f''  \cr &= - \frac{f}{z^2(f')}\left[\frac{z^4(f')^2}{f^3}\right]' \ge 0 
\end{align}
which shows that there exists a ``c-function" ($\sim -\frac{z^4(f')^2}{f^3}$) which is monotonically decreasing along this particular holographic renormalization group flow (when $f'(z)$ is positive).

This ``c-function" vanishes at the ``fixed point" of the renormalization group flow where $f'(z) = 0$. The inequality then demands that between the fixed points, $f(z)$  actually remains constant so that there is no flow between them at all. This example tells us that c-function so defined may be trivial, but it still constrains possible renormalization group flows.

One may wonder what will happen in the $AdS_3$ limit, where we apparently have stronger constraint from the null energy condition and the holographic c-function. In the limit, the space of the null vector relevant for the constraint opens up an extra dimension, and this gives a more stringent constraint on the renormalization group flow.

For more generic flows, apart from the issue of the integrability, we may not expect the existence of the c-function. The generic metric ansatz \eqref{ans} contains the Lifshitz-like scaling solution as well as other more exotic scaling solutions. It is rather surprising if we can construct the c-function that governs  the flows among all of them.
 Since the chiral version of the c-theorem from the field theory is not known, it would be important to investigate the structure of the holographic renormalization group flow in further details. The existence of the cyclic behavior  in some non-relativistic systems (e.g. \cite{Efimov}) may suggest that the holographic c-function may not non-trivially exist. We leave the question for future studies.

\section{Summary and discussions}

We have provided a gravity counterpart of the theorem by Hofman and Strominger that in $(1+1)$ dimension, chiral scale invariance indicates chiral conformal invariance. We have shown that the strict null energy condition gives a sufficient condition to guarantee the symmetry enhancement. The strict null energy condition can be understood as the unitarity condition on the dual $(1+1)$ dimensional quantum field theories.

In this paper, we have mainly focused on the local geometry, and did not address the problem of global structure. 
For instance, we could violate the invariance under the special conformal transformation by imposing non-trivial boundary conditions (see e.g. \cite{arXiv:1002.0615}\cite{arXiv:1004.3752} for related studies in the Schr\"odinger holography). It would be interesting to see what these global issues will tell us in the  dual field theory language. Relatedly, we would like to have a  better understanding on the nature of the microscopic causality or some other mechanism in non-relativistic theories that allowed us to use the Reeh-Schlieder theorem which played a crucial role in symmetry enhancement presented by Hofman and Strominger from the holographic viewpoint.

Generalizations to higher derivative gravity may not be difficult, but slightly non-trivial. We would like to set up the generalized version of the strict null energy condition along the line of reasoning in \cite{Nakayama:2010wx}, but since the geometry is not maximally symmetric, it would not be automatic that we could find the null vector which saturates the generalized energy condition. Moreover, it would  not be guaranteed that the left hand side of the equations motion is  semi-positive definite in the warped $AdS_3$ space once contracted with the null vectors. The latter may give a non-trivial constraint on the possible higher derivative corrections in gravity theories with good holographic interpretations.

Finally, in the null warped $AdS_3$ limit, the geometry can be regarded as the zero-dimensional version of the Schr\"odinger geometry \cite{Son:2008ye}\cite{Balasubramanian:2008dm}. It is an open question whether the Galilean invariant field theories with non-relativistic scale invariance would show the enhanced non-relativistic conformal invariance \cite{arXiv:0906.4112}, or the associated c-function exists. It would be challenging to attack this problem from the holographic perspective (see also \cite{arXiv:1011.5819} for a study of the holographic c-function in a similar setup).

\section*{Acknowledgments}
The author would like to thank the Michigan Center for Theoretical Physics for their hospitality where the initial stage of the work was done. He in particular would like to thank Cindy Keeler for stimulating discussions there. He also thanks fruitful discussions with D.~Honda and M.~Nakamura, who independently studied the same problem and concluded that the geometry must be warped $AdS_3$ \cite{HN}.
This work is supported  by the 
World Premier International Research Center Initiative of MEXT of
Japan. 


\end{document}